\documentclass[twocolumn]{aastex631}

\usepackage[utf8]{inputenc}
\usepackage{amsmath}
\usepackage{graphicx}
\usepackage{comment}

\begin{document}
\title{Phase and morphology of water-ice grains formed \\ in a cryogenic laboratory plasma}

\author{Andr\'e Nicolov}
\affiliation{Department of Applied Physics and Materials Science, California Institute of Technology \\
1200 E California Blvd, MC 128-95, Pasadena, CA 91125, USA}

\author{Murthy S. Gudipati}
\affiliation{Science Division, Jet Propulsion Laboratory, California Institute of Technology \\
4800 Oak Grove Drive, Mail Stop 183-301, Pasadena, CA 91109, USA}

\author{Paul M. Bellan}
\affiliation{Department of Applied Physics and Materials Science, California Institute of Technology \\
1200 E California Blvd, MC 128-95, Pasadena, CA 91125, USA}

\begin{abstract}
Grains of ice are formed spontaneously when water vapor is injected into a weakly-ionized laboratory plasma in which the background gas has been cooled to cryogenic temperatures comparable to those of deep space. These ice grains are levitated indefinitely within the plasma so that their time evolution can be observed under free-floating conditions. Using microscope imaging, ice grains are shown to have a spindle-like fractal structure and grow over time. Both crystalline and amorphous phases of ice are observed using Fourier Transform Infrared (FTIR) spectroscopy. A mix of crystalline and amorphous grains coexist under certain thermal conditions and a linear mixing model is used on the ice absorption band surrounding 3.2 $\mu$m to examine the ice phase composition and its temporal stability. The extinction spectrum is also affected by inelastic scattering as grains grow, and characteristic grain radii are obtained from Mie scattering theory and compared to size measurements from direct imaging. Observations are used to compare possible ice nucleation mechanisms, and it is concluded that nucleation is likely catalyzed by ions, as ice does not nucleate in absence of plasma and impurities are not detected. Ice grain properties and infrared extinction spectra show similarity to observations of some astrophysical ices observed in protoplanetary disks, implying that the fractal morphology of the ice and observed processes of homogeneous ice nucleation could occur as well in such astrophysical environments with weakly-ionized conditions.
\end{abstract}
                                         
\section{Introduction}
Grains of water ice and dust are observed in many astrophysical environments, such as protoplanetary disks \citep{terada2007} and interstellar clouds \citep{zubko2004}. These ice grains are formed from vapor at ultra-low pressures and temperatures, and typically exist within an environment that is weakly-ionized by stellar radiation or cosmic rays \citep{ivlev2015,zhang2022neutral}. Grains acquire electric charge from this ambient plasma, as first noted by \citet{spitzer1941dynamics}, because electrons have a higher collision frequency with the grains than do ions and so are collected by grains more frequently. The charge can affect ice grain formation \citep{bellan2022}, agglomeration \citep{ivlev2015}, accretion of water molecules \citep{bellan2020}, and dynamics \citep{Mitchell2006saturn, mendis2013comets}. Ice grain nucleation has traditionally been assumed to be heterogeneous, i.e., the grain forms as an ice coating on a non-ice nucleus which is typically carbon or silicate. However, observations of accretion disks \citep{potapov2021}, Saturn's rings \citep{hsu2018situ}, and aerosols in the Earth's atmosphere \citep{rapp2006modeling} suggest that in certain situations the ice grains are composed of pure water, rather than a surface coating of ice on a substrate. The formation of pure-water ice grains corresponds to homogeneous nucleation and is not well understood.

The ice phase depends on the formation conditions. Stable phases of water ice are the Ih phase (above $\sim72$ K) and the XI phase (below $\sim72$ K). These phases do not form instantaneously. Instead, below $\sim130$ K ice typically forms in a metastable amorphous phase since accreting water molecules do not have the energy to orient themselves into a crystalline structure. Consequently, it takes from minutes to billions of years for amorphous ice to transition to the stable crystalline phase, depending on the thermal environment \citep{mckinnon2005, mastrapa2013amorphous}. Both amorphous and crystalline phases have been detected in the solar system and in deep space \citep{mastrapa2013amorphous}. The ice phase is important to the surface chemistry of the ice grain and its interaction with high-energy particles and radiation common in space plasma environments \citep{lignell2015, gudipati2023thermal}.

Astrophysical ice analogs can be made in the laboratory and compared to astrophysical measurements. These analogs typically reproduce some, but not all, features of astrophysical ice. A standard laboratory method is to deposit thin films of water ice on a substrate. Spectral techniques are then used to study formation of amorphous and crystalline phases \citep{mastrapa2013amorphous}, the absorptive and refractive behavior at varying temperatures \citep{gudipati2023thermal}, and the ice interaction with high-energy radiation occurring in astrophysical environments \citep{allodi2013complementary}. Laboratory-derived optical constants can be used to calculate the transmission spectra of astrophysical ice dusts and aerosols and interpret telescope measurements of these astrophysical phenomena.

The formation of various types of dust grains has been studied within laboratory plasmas, which simulate the free-floating conditions and ion chemistry present in many astrophysical settings, as grains are formed within the plasma bulk and are electrostatically levitated \citep{bouchoule1999dusty}. Previous studies have focused on chemical precipitation within weakly-ionized plasmas of solid aerosols, such as analogs of Titan's tholins \citep{szopa2006pampre} and polycyclic aromatic hydrocarbons that are common in the interstellar medium \citep{ricketts2011cosmic}. One common configuration, used by \citet{szopa2006pampre}, is the steady-state radio-frequency (RF) capacitively-coupled plasma, in which a sustained plasma discharge is produced between parallel-plate electrodes. The plasma develops a positive electric potential, while grains acquire a negative charge \citep{shukla_mamun_2002}, causing the grains to be electrostatically suspended within the plasma while they accrete new material and interact with their surroundings both physically and chemically. The behavior of these capacitively-coupled RF plasmas is well-studied, providing a reliable and fully-characterized plasma environment in which to study grain formation.

RF plasmas have recently been utilized for studying ice grains by arranging the system to have a sufficiently cold background gas temperature. \citet{Shimizu2010} first succeeded in forming a cloud of water-ice grains within a plasma by cooling the RF electrodes in a liquid nitrogen bath. A background gas composed of deuterium and oxygen was cooled through contact with the cold electrodes. Upon being weakly-ionized, the gas reacted to form supersaturated D${_2}$O vapor from which ice grains nucleated. An experiment by \citet{chai2015} used a similar liquid nitrogen-cooled RF plasma, but directly injected water vapor into an argon background gas, and observed nucleation and then growth of ice grains up to 700 microns in length. Measurements were made of the infrared absorption spectra of these ice grains. This approach gave control of the background gas composition and amount of water vapor, and also allowed formation of ice grains of other molecules, such as methanol and acetone.

Both these systems were limited by the non-adjustable temperature provided by liquid nitrogen cooling of the electrodes. This prevented these studies from determining a temperature dependence of the infrared absorption spectra of the ice grains. Furthermore, the measured gas temperatures reached in these setups were relatively high ($>170$ K according to \citet{Marshall2020}), which prevented access to the low-temperature regime where amorphous ice forms. We report here initial results from a new apparatus where the temperature is adjustable and can be much lower than previously attained. This paper presents observations over a range of cryogenic temperatures of the formation, growth, and phase composition of ice grains formed in this ultra-cold plasma. 

\section{Experimental setup}
\begin{figure}[h!]
  \centering
  \includegraphics[width=8.5cm]{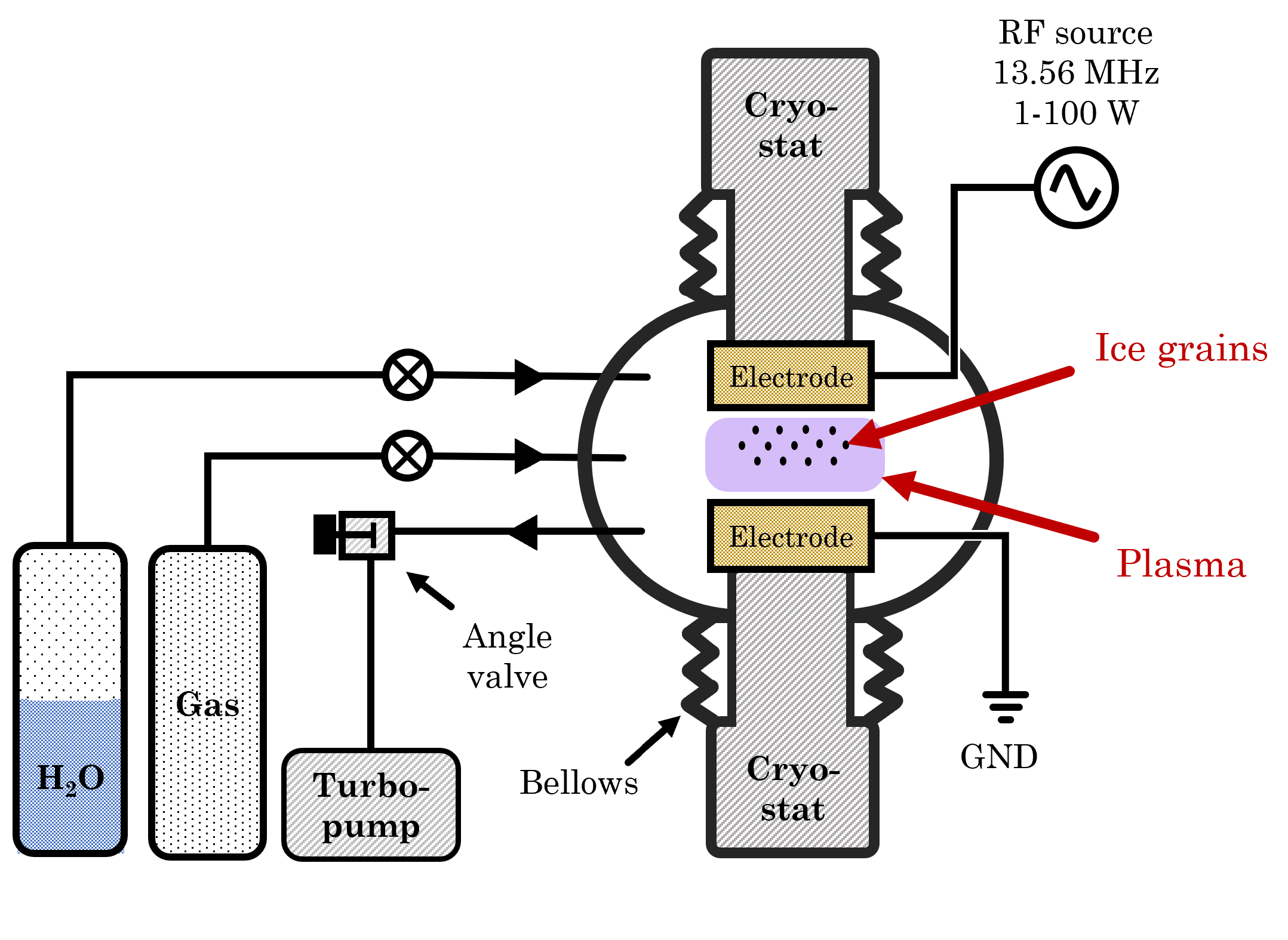}
  \caption{Schematic of the Ice Dusty Plasma Experiment at Caltech.}
  \label{schem}
 \end{figure}

The new experiment was constructed at Caltech to produce astrophysical ice analogs within a controlled low-temperature plasma environment. The apparatus contains a vacuum system to maintain low pressure, cryogenic cooling for temperature control, and a radio-frequency plasma generator to maintain ionization of a background gas; a schematic is shown in Figure \ref{schem}. This apparatus is similar to those used to make ice-dusty plasmas by \citet{Shimizu2010} and by \citet{chai2015}, but is capable of precise control of electrode temperatures and can reach much colder conditions. 

Parallel-plate copper electrodes, spaced 1-5 cm apart (adjustable under vacuum), are each mounted on a cold head of a closed-cycle liquid helium cryogenic compressor (SHI Cryogenics). Temperature-sensitive silicon diodes embedded in the electrodes measure their temperature, and embedded heater cartridges maintain each electrode at a desired preset temperature via a feedback system (Stanford Research Systems). The electrodes are capable of maintaining steady temperatures between 50 and 200 K. 

The chamber is pressurized to between 100 and 2000 mTorr (1 Torr $\sim 1.33$ mbar) by a steady flow of background gas (typically hydrogen or argon) through an adjustable leak valve. The pumping speed is adjusted using an angle valve between the chamber and a turbopump. This causes a very small flow ($<10$ SCCM) in and out of the 3.8-liter chamber, such that the gas remains in the chamber long enough to come to a conductive thermal equilibrium profile \citep{chai2018}. While the gas nearest to the chamber walls remains near room temperature, the gas in the inter-electrode gap maintains an average temperature closer to the cryogenic temperature of the electrodes. In a similar experiment by \citet{Marshall2020}, in which electrodes were cooled using liquid nitrogen, the average temperature of the gas between the electrodes was found to be 30-40 K higher than the electrode temperature; the temperature difference in this experiment is comparable. When electrode temperatures are quoted throughout this paper, it must be understood that the gas (and ice) temperature is some tens of kelvins higher.

An RF power source with an automatic impedance matching network (T\&C Power Conversion) applies a 13.56 MHz RF voltage (1-100 W) across the electrodes, breaking down the gas to produce weakly-ionized plasma (ionization fraction of about $10^{-6}$) in the inter-electrode gap. Water vapor is then injected through a second leak valve and rapidly cools upon mixing with the cold background gas. The vapor is supersaturated upon entering the chamber due to the low gas pressure and temperature, and ice grains form spontaneously from the vapor within the plasma; no ice grains form in the absence of the plasma. The grains quickly acquire a negative charge through electron collection and are perfectly trapped in the positive potential of the plasma. The resulting cloud of ice is thus confined to a region between the electrodes where temperature is sufficiently low and ionization is sufficiently high. The setup allows direct control of pressure, gas flow rate, water vapor flow rate, and the temperature of each electrode. Electric field strength and ionization fraction can be indirectly adjusted by changing the input RF power and the electrode spacing.

\section{Diagnostics and configuration}
 \begin{figure*}
  \centering
  \includegraphics[width=16cm]{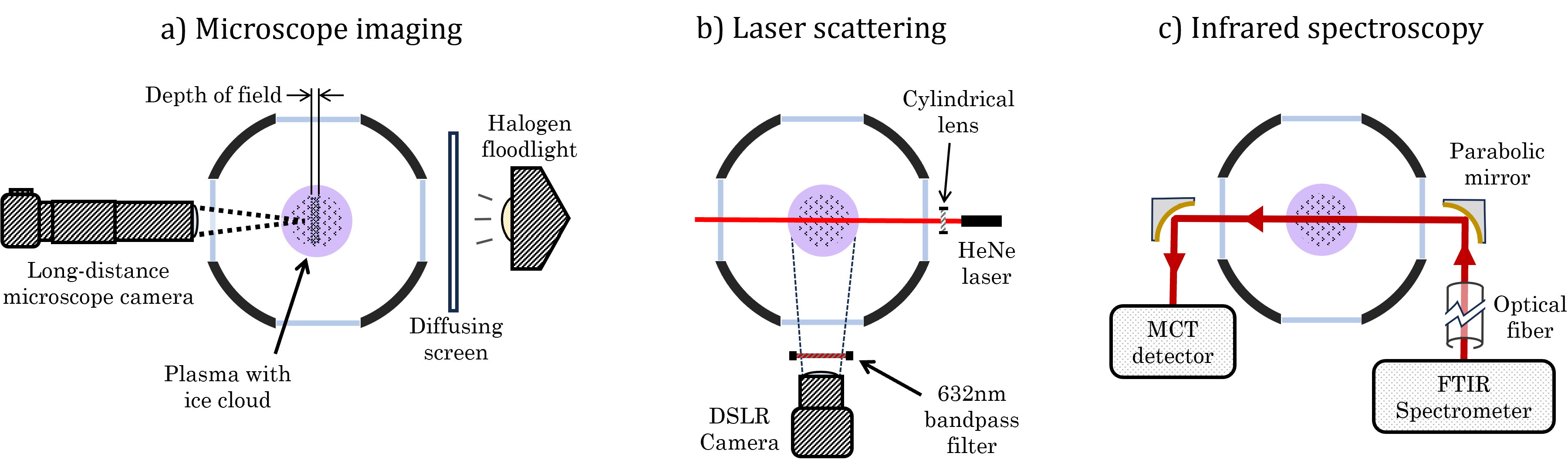}
  \caption{Schematic sketches of diagnostics (top view). (a)  A long-distance microscope camera images ice grains within a $\sim 30$-micron depth of field, which are back-lit by a bright halogen lamp. (b) The cloud of ice grains is illuminated by a sheet of laser light and photographed using a DSLR camera. (c) IR beam is emitted by an FTIR spectrometer, is transmitted through the plasma, and is collected by an MCT detector.}
  \label{diagnostics}
 \end{figure*}
 
The infrared extinction spectrum of the ice grains is measured using a Thermo-Nicolet iS50 Fourier Transform Infrared (FTIR) spectrometer. A modulated infrared (IR) beam emitted by the FTIR spectrometer is focused by a gold parabolic mirror to enter a 1.5 meter long IR optical fiber (Guiding Photonics), and upon exiting the fiber, is focused into a beam by another gold parabolic mirror. This beam passes through a sapphire window into the vacuum chamber, traverses the plasma and ice cloud, exits the vacuum chamber through another sapphire window, and focuses onto a mercuric cadmium telluride (MCT) detector (Figure \ref{diagnostics} (c)). The beam is partially absorbed at some wavelengths by ice grains in the beam path, and the spectrometer provides scans of the ice extinction spectrum. Each scan involves 60 separate spectra that are averaged to increase the signal-to-noise ratio, and takes roughly one minute. 

\begin{figure}
       \centering
    \includegraphics[width=8.5cm]{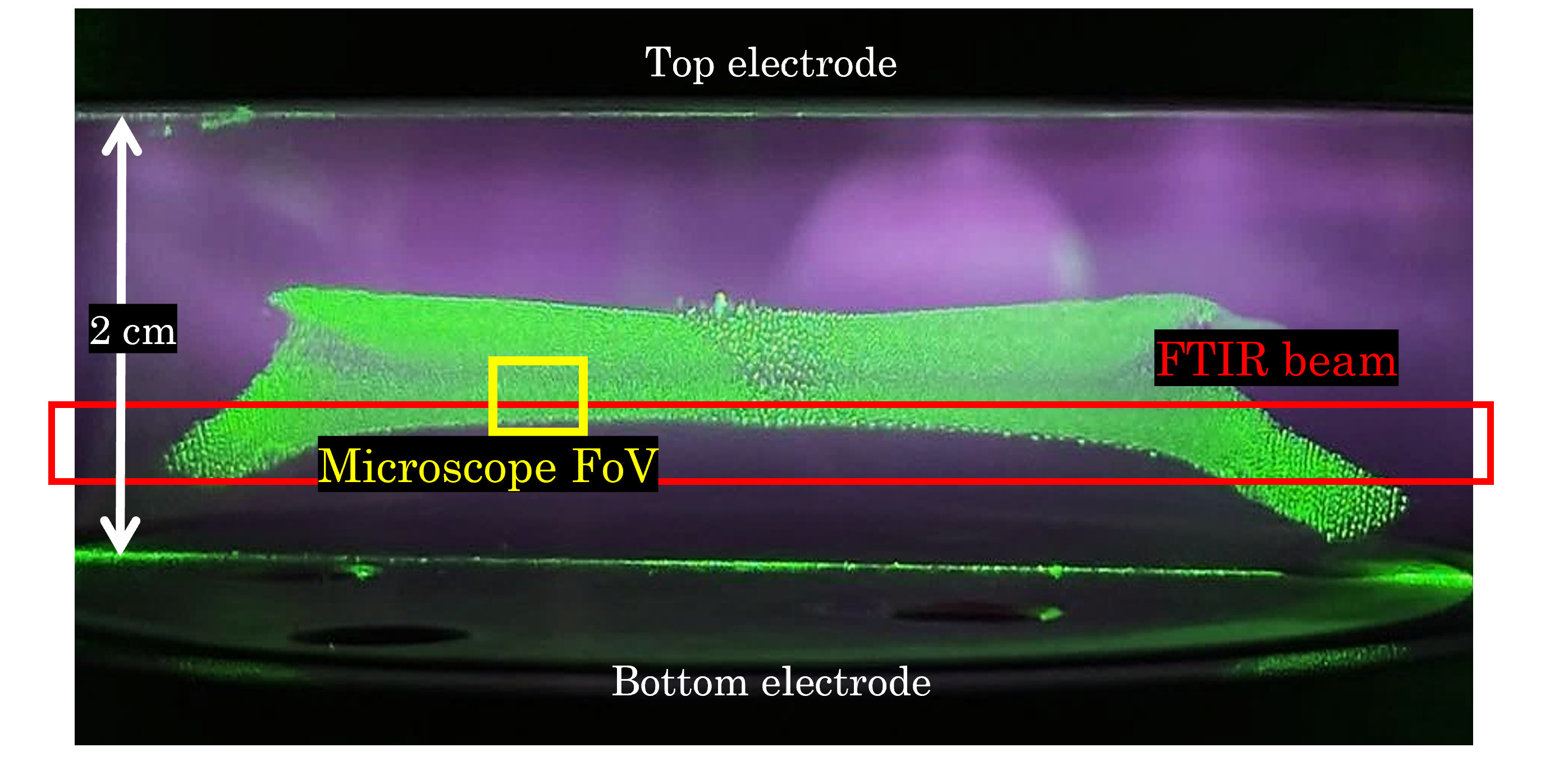}
    \caption{Photo of the icy plasma; hydrogen plasma (purple) contains laser-illuminated water-ice grains (green). FTIR beam path is outlined in red, and approximate microscope camera field of view is outlined in yellow.}  
    \label{laser scattering}
\end{figure}

Ice grains in the experiment are visualized using two different imaging techniques. In the first scheme, a long-distance microscope lens (Infinity Photo-Optical) with a shallow depth of field ($\sim 30 - 100$ $\mu$m), mounted to a Nikon DSLR camera, captures silhouettes of ice grains backlit by a halogen flood lamp, as shown in Figure \ref{diagnostics} (a). The field of view is approximately 5 mm by 5 mm, with a resolution of 3 microns. A second imaging scheme utilizes a HeNe laser sheet, created by a cylindrical lens, to illuminate a cross-section of the ice cloud. Scattered light, filtered through a 632.8 nm bandpass filter, is captured by a DSLR camera with a wide-angle macro lens fixed with a tube extender. This is shown in Figure \ref{diagnostics} (b). This diagnostic shows the shape of the cloud, monitors how much ice is in the FTIR beam path, and reveals large-scale dynamics of collections of ice grains. Figure \ref{laser scattering} shows the plasma and light scattered off the ice cloud, in which a green laser pointer is used in place of the HeNe laser for enhanced color contrast between ice grains (green) and the plasma (purple). The regions intercepted by the IR beam and the microscope camera are indicated.

\begin{figure}
       \centering
    \includegraphics[width=8.5cm]{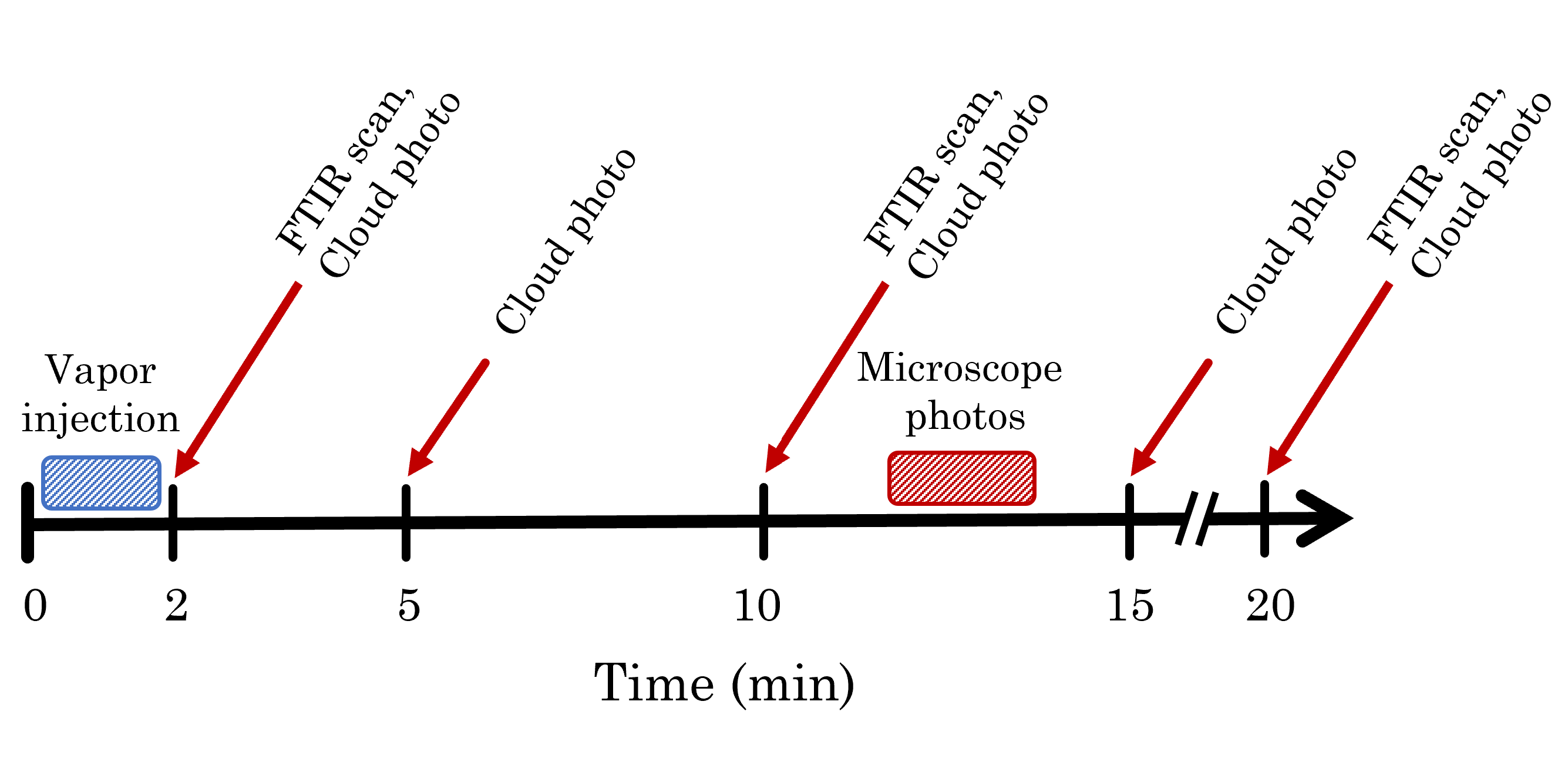}
    \caption{Experimental timeline. In the first two minutes, a flow of vapor is introduced to the experiment. FTIR scans and wide-angle laser scattering photos of the ice cloud are taken at $t = 2$, 10, and 20 minutes; cloud photos are also taken at 5 and 15 minutes. Microscope photos are taken between 10 and 15 minutes.}  
    \label{procedure}
\end{figure}

Data was collected using the following procedure. The electrodes were first brought to temperature; tested temperatures ranged from 80 to 150 K, in 10 K increments. RF power and electrode separation were fixed at 75W and 2 cm, respectively. The chamber was filled with hydrogen gas and ionized by the RF source. Background gas pressures of 400, 800, and 1600 mTorr were each tested. After the RF source generated a plasma, water vapor was introduced into the chamber from a nozzle located about 3 cm from the outer edge of the electrodes. The water vapor was injected into the chamber at approximately 1 SCCM for two minutes as ice grains spontaneously formed within the plasma. At $t = 2$ minutes, the water flow was shut off to stop new ice grain nucleation. FTIR scans were taken at 2 minutes (vapor shutoff), 10 minutes, and 20 minutes after vapor injection began ($t = 0$). These scans were accompanied by laser scattering photos to show the ice cloud position and structure, taken at $t = 2$, 5, 10, 15, and 20 minutes. In addition, long-distance microscope photos showing the size and shape of individual ice grains were taken between the 10- and 15-minute mark; the ice grains were typically too small to image directly before $t = 10$ minutes. A timeline of each experimental run is shown in Figure \ref{procedure}.

\section{Ice formation and growth with pressure and temperature}
\begin{figure*}
  \centering
  \includegraphics[width=15cm]{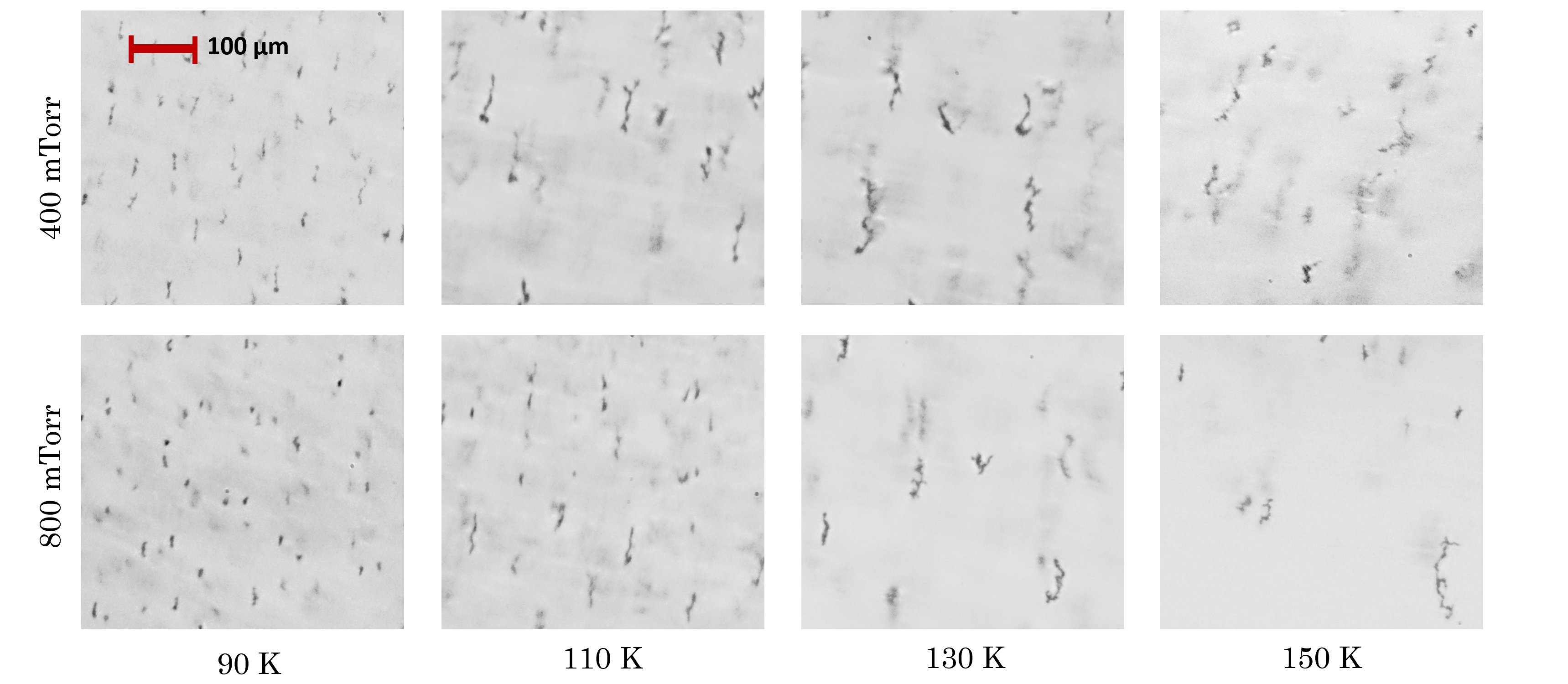}
  \caption{Microscope photos of ice grains. Four temperatures and two pressures are shown; at 1600 mTorr, grains were smaller than the microscope resolution, and photos at intermediate temperatures are available through a data repository at doi:10.22002/x3qp1-3k378.} Grains within the 30-micron depth of field of the microscope appear dark against a white background, which is unfocused light from the floodlight behind the chamber (see Fig. \ref{diagnostics} (a))
  \label{grain photos}
 \end{figure*}

The long-distance microscope images in Figure \ref{grain photos} show ice grains with a spindle-like fractal shape where the spindles generally align in the vertical direction. This is similar to ice grains observed in previous experiments \citep{chai2015, Marshall2017}. These photos show that grains were larger at lower pressures, consistent with observations by \citet{chai2015}. Grains were generally larger at warmer conditions, a new result enabled by the ability to control temperature. Smaller grains tend to be more closely-packed. This dependence indicates that grain size increases with the inverse of the gas density ($n \sim p/T$). This dependence is somewhat surprising: because a lower neutral density means fewer water molecules are contained in the plasma at any time (the ratio of water vapor to H$_2$ pressure was the same for all configurations), one might expect that a higher density would lead to more ice accretion, but the opposite is observed. Because no new water vapor was introduced to the chamber after $t=2$ minutes, one might also expect the size to depend directly on the sublimation rate; however, as sublimation increases with pressure and temperature, and grain size increases with temperature but with the inverse of pressure, grain size cannot be treated as a simple function of the sublimation rate. These observations suggest a more complex process determining particle growth rate, perhaps dependent on vapor diffusion speeds, which will require further experimental study outside the scope of this paper. While the size of an ice grain is influenced by the environment, the grain shape and structure do not appear to change, even when the FTIR spectra indicate a phase transition from amorphous to crystalline ice (discussed later). The elongated fractal structure of the ice grains thus appears to be linked to how the ice grain grows within a plasma and not to the internal microstructure of the ice.

\begin{figure}
  \centering
  \includegraphics[width=8.5cm]{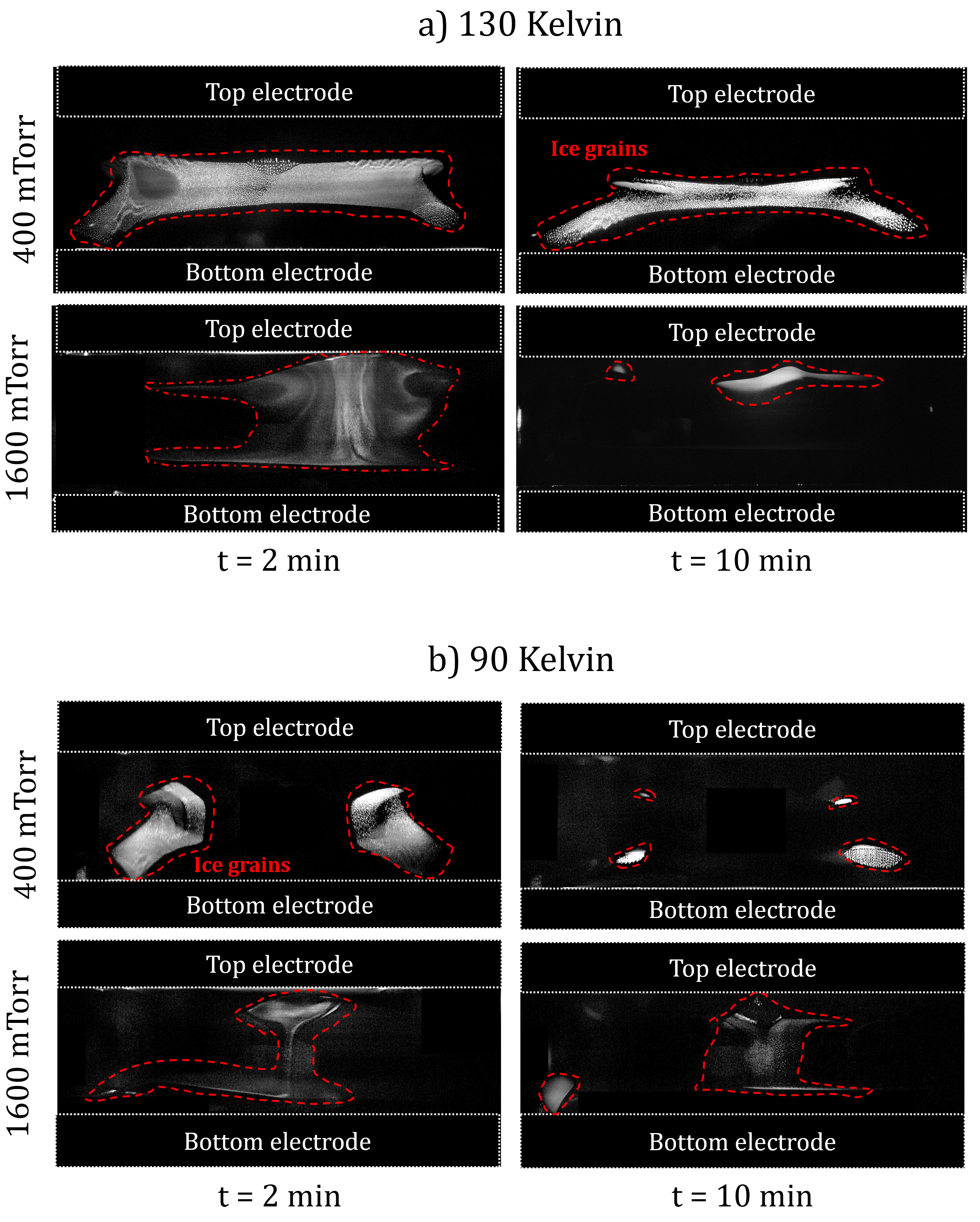}
  \caption{Wide angle laser scattering photos of the ice cloud, at two electrode temperatures, two pressures, and two points in time. The bright regions are the scattered light off the ice grains; electrodes are at the top and bottom of each photo.}
  \label{cloud photos}
 \end{figure}

Laser scattering photos of the ice cloud are shown in Figure \ref{cloud photos} at 400 and 1600 mTorr, $t = 2$ and 10 minutes after ice formation. Figure \ref{cloud photos} (a) and Figure \ref{cloud photos} (b) show two different temperatures. The complex shape of the ice cloud is caused by the intricate dynamics of grains within the plasma; these dynamics are influenced by grain size, pressure, and temperature \citep{shukla_mamun_2002}. Grains are observed to be in constant motion within the cloud, showing vortices that transport grains to and from the center of the plasma. The laser scattering consistently gets brighter over time in the experiments at 400 and 800 mTorr, indicating that there is ongoing grain growth, as larger grains scatter more light. Conversely, at 1600 mTorr, the size and brightness of the ice cloud decreases significantly by 10 minutes, likely due to increased sublimation at high pressure.
At all pressures, more ice generally appears at 130 K than at 90 K, suggesting less efficient nucleation at colder temperatures. This temperature trend persists across the entire 80 K to 150 K temperature range, with higher temperatures leading to larger and brighter ice clouds.

\begin{figure}[h]
  \centering
  \includegraphics[width=8.5cm]{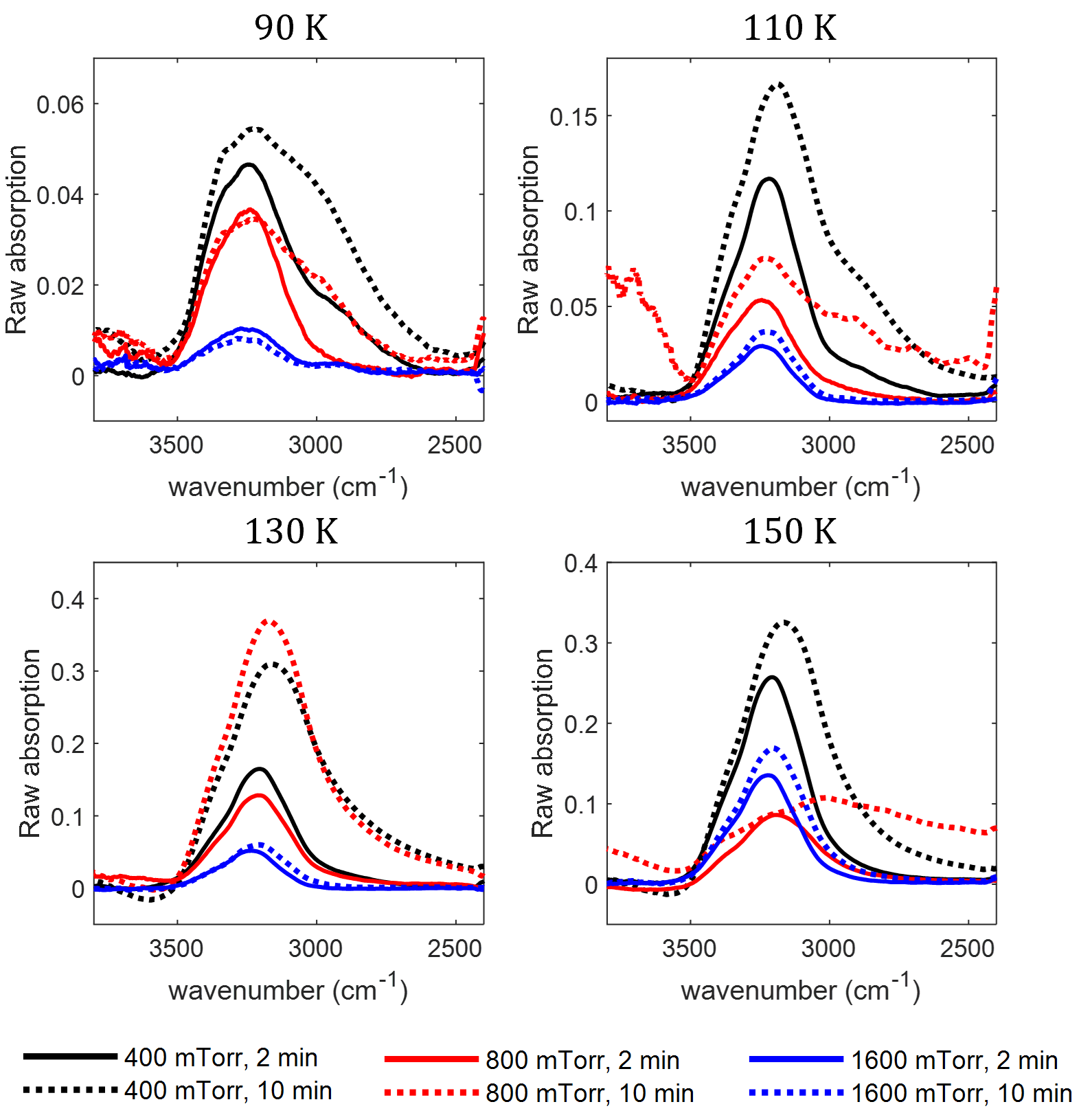}
  \caption{Spectra taken at 90, 110, 130, and 150 K. Raw spectra at all temperatures will be made available through the Caltech data repository\textbf{, doi:10.22002/x3qp1-3k378}. All three pressures tested are shown. Solid-line spectra are taken at $t = 2$ min, and dotted-line spectra at $t = 10$ min.}
  \label{magnitudes}
 \end{figure}

Extinction spectra measured by the FTIR spectrometer in the 800-4200 cm$^{-1}$ range reveal strong absorption by water ice in the O-H region near 3200 cm$^{-1}$ ($\sim3$ microns); this is shown in Figure \ref{magnitudes}. The magnitude and shape of the extinction curves vary with the temperature, pressure, and time since ice formation. The magnitude varies with the quantity of ice that intercepts the IR beam, accounting for both number of grains and grain size. The ice quantity is sensitive both to changes in the total amount of ice deposited in the plasma and to movement of the ice cloud in or out of the IR beam path. Based on the position of the ice in the laser scattering photos (see Figure \ref{laser scattering}), the effect of movement is generally negligible.

The magnitude of the absorption spectrum is larger when the electrode temperature is higher: At 90 K, the absorption magnitude is 3-5 times smaller than the absorption at 130 K. Similarly, the absorption magnitudes are considerably smaller at high pressures than at low pressures. The absorption magnitude therefore correlates with the grain size: when grains are small, infrared absorption through the ice cloud is low. This shows that when grains are smaller, less total ice is contained in the plasma; the small grain size is not countered by higher numbers of ice grains. This again indicates that nucleation is less efficient at lower temperatures, as less of the water vapor injected into the chamber becomes ice detected by the FTIR spectrometer.

At 90 K, the extinction magnitudes do not change much between the 2- and 10-minute scans. At 130 K in contrast, for both the 400 and 800 mTorr configurations the signal grows significantly over time. As no new vapor was introduced at these times, the increasing amount of ice must be entirely due to accretion of vapor that sublimated from ice outside the FTIR beam path, either from ice frozen to the electrodes or from other ice grains not in the IR beam path. Because sublimation happens more readily at higher temperatures, grains at 130 K are immersed in a higher partial pressure of water vapor than grains at 90 K, and can continue to grow through accretion, while grains at 90 K do not grow once the initial vapor is depleted.

\section{Scattering and grain growth}
\begin{figure}[h!]
  \centering
  \includegraphics[width=8.5cm]{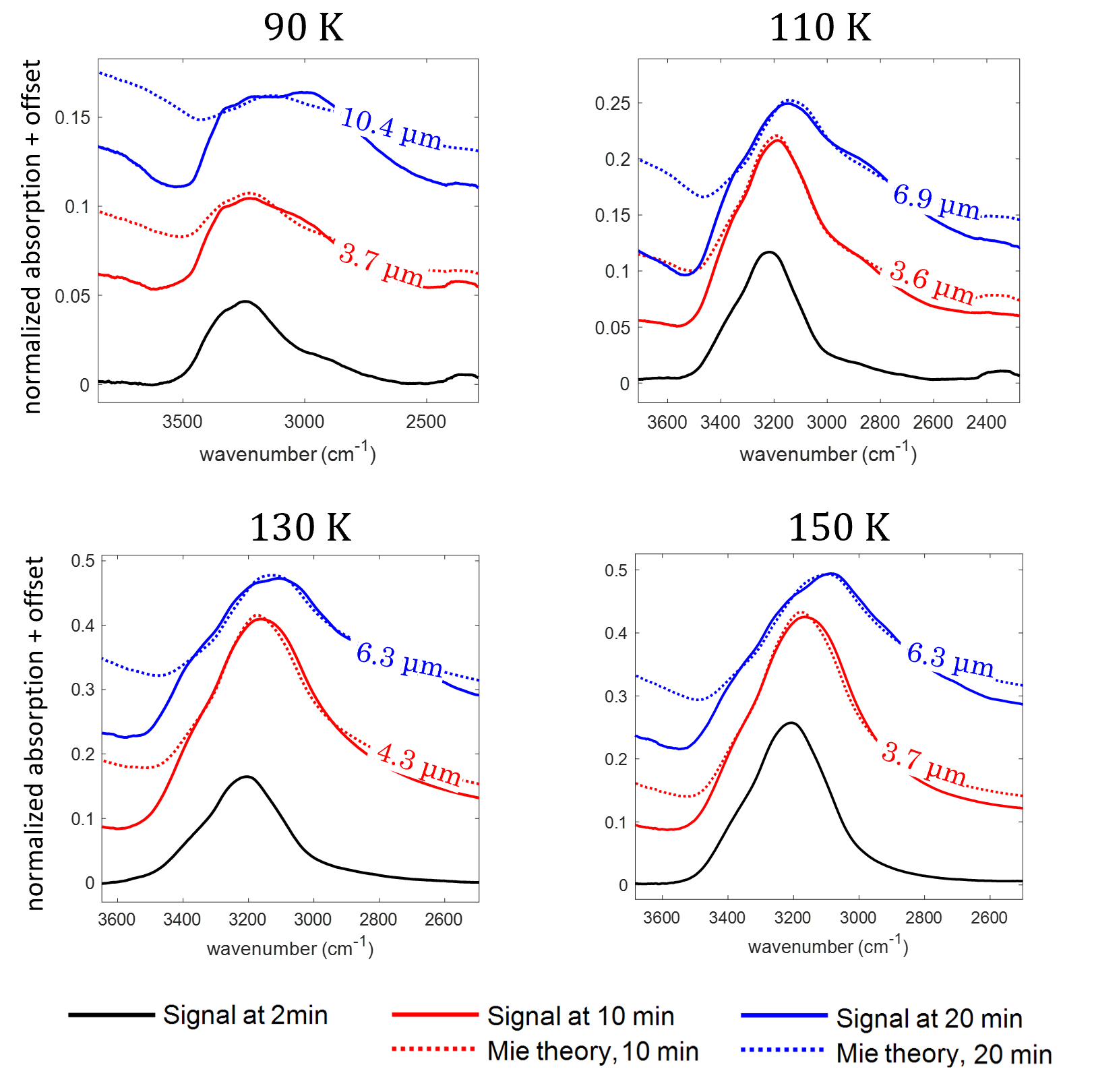}
  \caption{Spectra obtained at $t = 2$, 10, and 20 minutes, and Mie scattering fits to the $t = 10$ and 20-minute spectra. Shown are spectra taken at 400 mTorr, and temperatures of 90, 110, 130, and 150 K. Curves are offset in the y-direction for clarity. Mie fit radii and uncertainties at all pressures and temperatures are presented in Fig. \ref{grain sizes}.}
  \label{Mie fit}
 \end{figure}

The solid lines in Figure \ref{Mie fit} show normalized extinction spectra taken at 400 mTorr over a range of temperatures, at 2, 10, and 20 minutes after ice formation. At later times, the spectrum broadened and shifted due to inelastic scattering; the change is more pronounced at lower pressures and higher temperatures, correlating with grain growth.

A method developed by \citet{clapp1995} is used to estimate grain size from the change in the spectra. In this method, a spectrum that shows no scattering is used to provide the absorption coefficient $k(\omega)$ for the ice grains; here $\omega $ is the optical frequency. The index of refraction $n(\omega)$ is then calculated using a Kramers-Kronig transformation,
\begin{equation}
    n(\omega) = \frac{2}{\pi} (\omega^2 - \omega_0^2) P \int_0^\infty \frac{w k(\omega) dw}{(\omega^2-w^2)(\omega_0^2-w^2)} + n(\omega_0)
\end{equation}

\noindent where $P$ indicates taking the Cauchy principle value of the integral. The wavenumber $\omega_0$ is an ``anchor point" at which the index of refraction is known; here we use $\omega_0 = 4000$ cm$^{-1}$, at which $n(\omega_0) = 1.232$ \citep{clapp1995}. 

The optical constants $n(\omega)$ and $k(\omega)$ are inserted into a Mie scattering algorithm by \citet{bohren2008absorption} for scattering off a uniform sphere (BHMie), along with a guess for the particle radius; the algorithm then calculates an altered spectrum that includes scattering by an ice grain. The guess for the radius is iterated until the produced Mie spectrum best fits the observed spectrum. These fits are plotted in Figure \ref{Mie fit} (dotted lines) for the 400 mTorr spectra at a range of temperatures. These effective Mie radii range between 1 and 14 microns. Overall, this simple Mie scattering method well-replicates the observed time dependence of the spectrum. The model fits observations very well around the absorption band, but diverges on the high-wavenumber side when grains are large. This could be due to the distribution of grain sizes or to their non-spherical structure. 

\begin{figure}[h!]
  \centering
  \includegraphics[width=8.5cm]{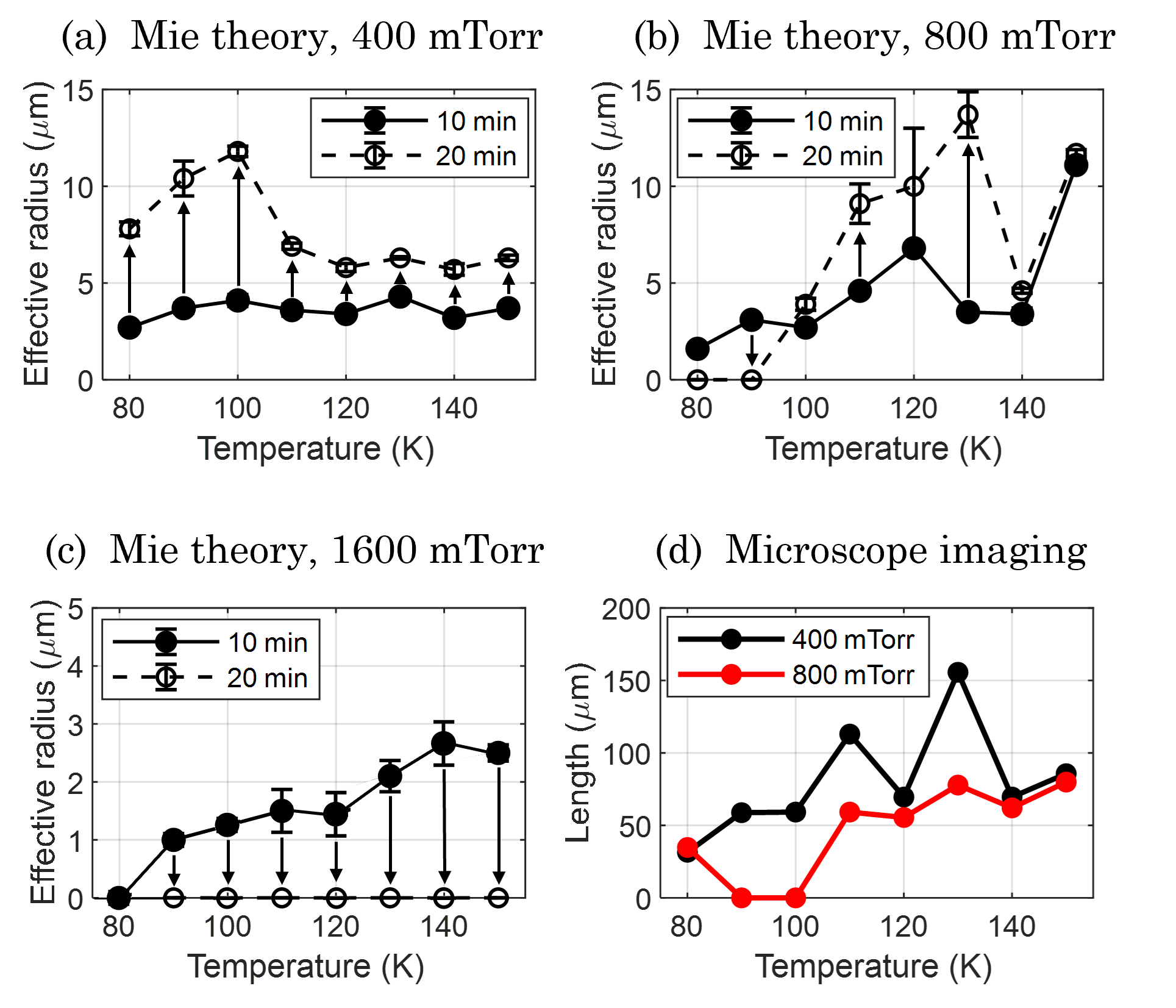}
  \caption{Trends of grain size with temperature and time. Graphs (a) through (c) show effective Mie scattering radii at different pressures; closed circles are radii at 10 minutes, and open circles are radii at 20 minutes. Arrows indicate size change over time. Graph (d) is the grain length from microscope photos, taken just after $t = 10$ minutes.}
  \label{grain sizes}
 \end{figure}

Figures \ref{grain sizes} (a) – (c) present effective radii from Mie scattering versus temperature. Generally, effective radii increase with temperature at 800 mTorr and 1600 mTorr, while this trend is not evident at 400 mTorr. At 400 and 800 mTorr, ice grains generally grew between the spectral scans at 10 and 20 minutes. At 1600 mTorr, ice vanished by the 20 minute mark due to more rapid sublimation. In Figure \ref{grain sizes} (d), grain lengths from microscope images taken just after $t = 10$ minutes are plotted against temperature. Similar to the Mie scattering results, grains are larger at higher temperatures at 800 mTorr, with a less clear trend at 400 mTorr. At 1600 mTorr, all grains within the field of view of the camera were smaller than the camera's resolution (< 3 $\mu $m), so no grain size could be determined.

\begin{figure}[h!]
  \centering
  \includegraphics[width=8.5cm]{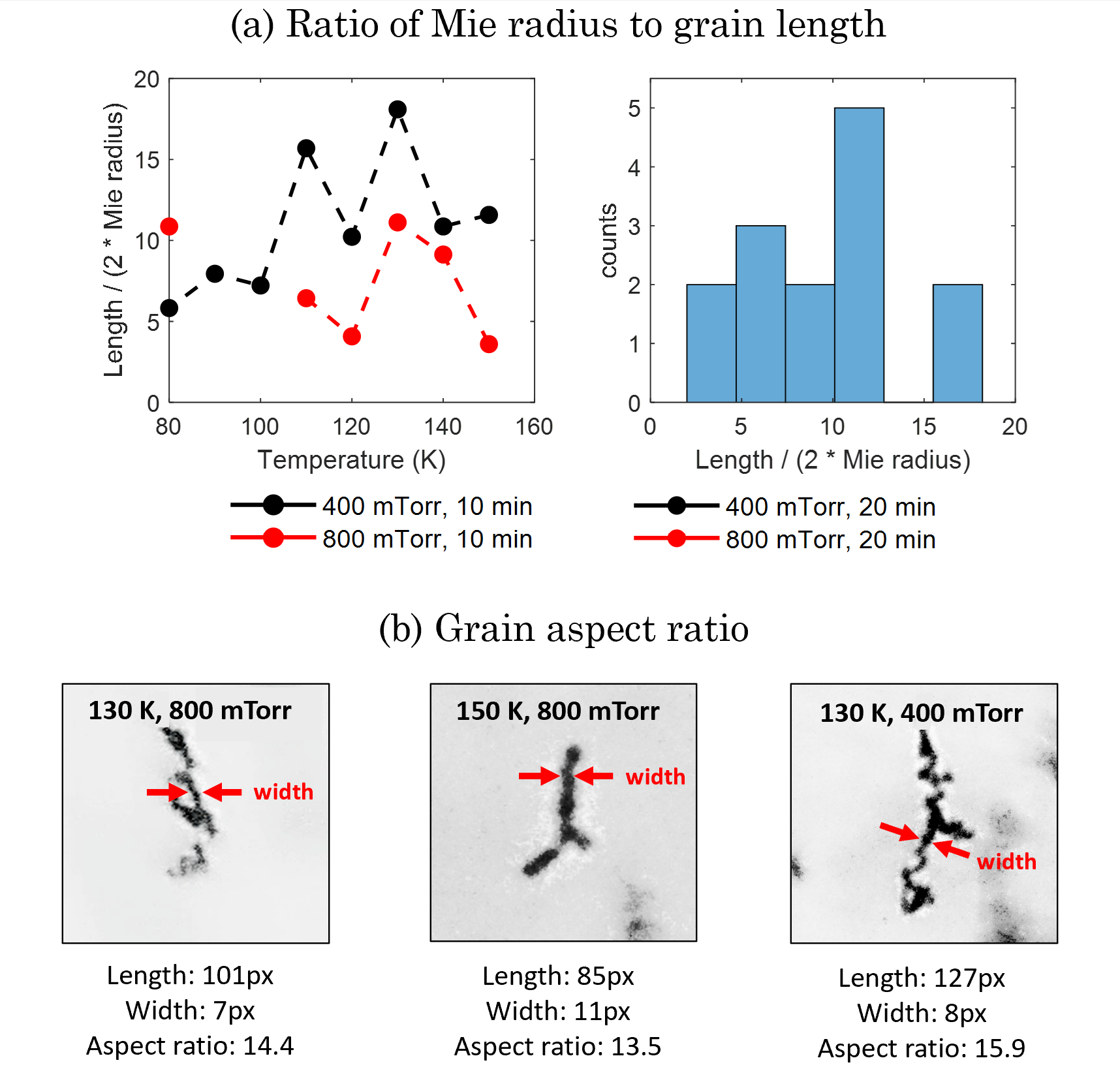}
  \caption{(a) Plot of ratios between grain length (microscope photos) and Mie scattering effective radii ($t = 10$ min) for 400 and 800 mTorr configurations. Plot at left shows variation with temperature, and plot at right shows a histogram of all data points. (b) Microscope photos of grains in which the width can be resolved. Grain width is measured across the thinnest part of the grain. Images are compressed to roughly 2 microns per pixel.}
  \label{aspect ratio}
 \end{figure}

An obvious issue is that there is a tenfold difference between measured grain lengths and Mie scattering-derived effective radii. The ratio of grain length to Mie scattering diameter is displayed in Figure \ref{aspect ratio} (a), both showing trends with pressure and temperature and a histogram of all values. This ratio fluctuates anywhere between 4 and 20, with a modal average value around 12. It does not have an obvious dependence on temperature.

This discrepancy is attributed to the elongated, fractal grain shape. At most polarization angles, infrared radiation intercepts a cross-section on the order of the width of the ice grain, rather than its total length. Because of the limited resolution of the long distance microscope photos, precision measurement of the width is not possible for most grains. Photos of some exceptionally large grains, in which their widths are within camera resolution, are shown in Figure \ref{aspect ratio} (b). These images exhibit aspect ratios ranging from 13 to 16. These are comparable in magnitude to the ratio of length to Mie scattering diameter shown in Figure \ref{aspect ratio} (a), providing some explanation for the discrepancy between microscope measurements and Mie theory calculations.

\section{Phase analysis of the ice grains}
\begin{figure}[h!]
  \centering
  \includegraphics[width=8.5cm]{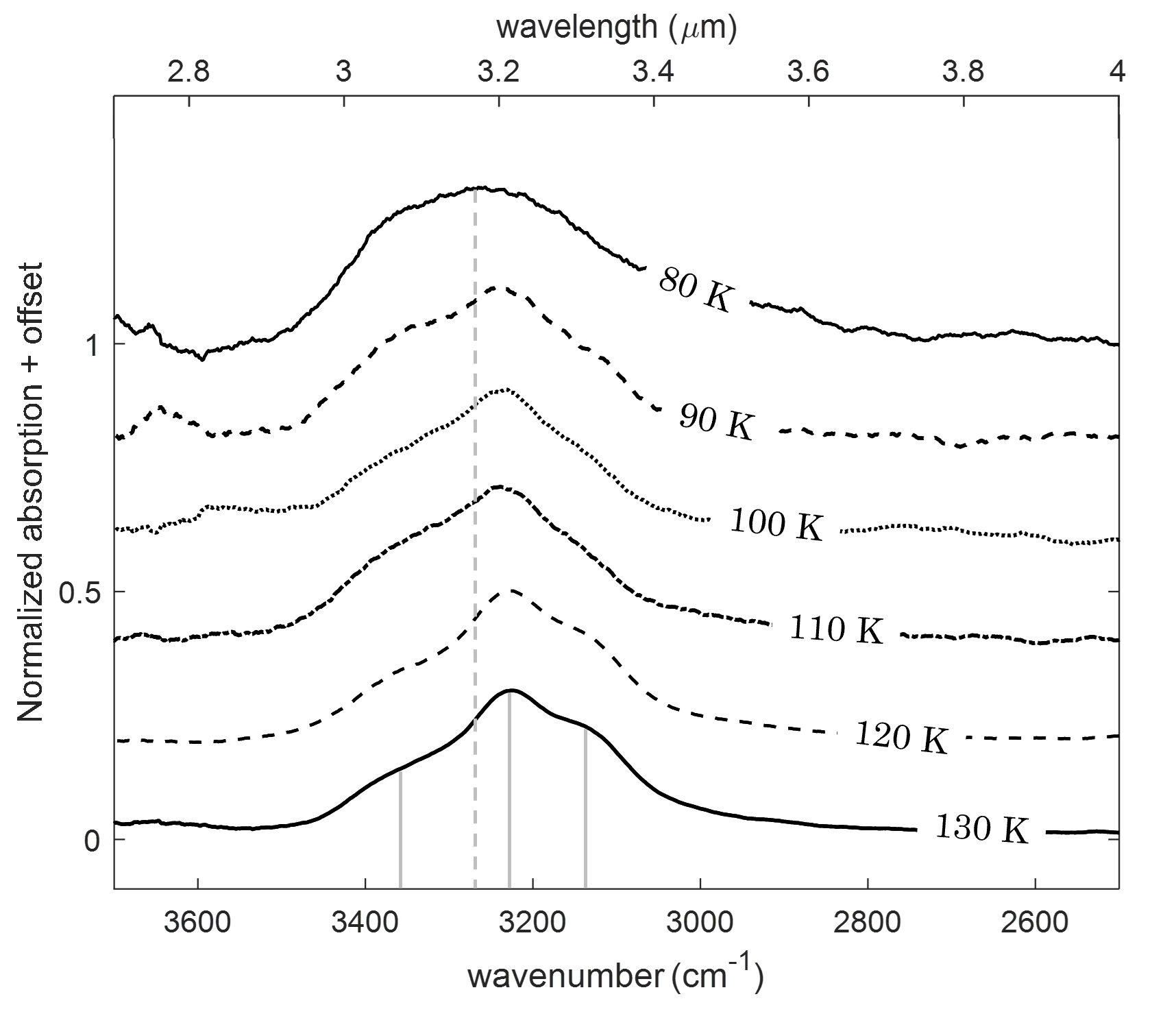}
  \caption{Ice spectra at formation, with electrode temperatures from 80 K to 130 K, all at t = 2 min and p = 800 mTorr. Spectra at 140 and 150 K are not shown here as they displayed the same crystalline characteristics to the scan at 130 K. Spectra are offset in the vertical direction by an arbitrary amount for clarity. Vertical grey lines indicate positions of peaks: the solid lines show the three crystalline peaks, and the dotted line shows the amorphous peak.}
  \label{spectrum at formation}
 \end{figure}

The shape of the infrared extinction band is determined both by inelastic scattering (due to ice size and shape) and by the ice phase (amorphous or crystalline). When grains are considerably smaller than the IR wavelengths (Rayleigh scattering regime), as is the case for the $t = 2$ minute FTIR scans, scattering does not affect the spectrum. Figure \ref{spectrum at formation} shows normalized spectra at electrode temperatures ranging from 80 K to 130 K; note that the gas and ice are about 20 to 30 K warmer than the electrodes.

At 130 K, the spectrum displays a trimodal shape with a maximum at 3215 cm$^{-1}$ and smaller humps at wavenumbers of 3360 and 3140 cm$^{-1}$. Peak positions are indicated in Fig. \ref{spectrum at formation} by solid grey vertical lines. This shape indicates a crystalline structure, specifically the Ih phase of water ice. At 80 K, the spectrum broadens to become unimodal, with its singular peak shifted to 3270 cm$^{-1}$ (vertical dashed line in Fig. \ref{spectrum at formation}); this indicates an amorphous ice phase consistent with observations of ice films formed at similar temperatures \citep{mastrapa2013amorphous}. The spectra from 90 to 110 K display a mix of both amorphous and crystalline properties: they contain a leftward-shifted broad spectrum, but retain a local peak at 3215 cm$^{-1}$. This indicates that both amorphous and crystalline ice are present within the cloud of ice grains: either individual grains contain both amorphous and crystalline ice domains, or some of the ice grains are crystalline and others are amorphous. The latter is more likely, as the thermal conductivity of crystalline ice is sufficiently high to keep the entire grain at a uniform temperature \citep{mastrapa2013amorphous}. Amorphous ice has a thermal conductivity orders of magnitude lower than crystalline ice, but the amorphous-to-crystalline phase transition is exothermic \citep{gudipati2023thermal} providing a runaway crystallization process in an ice grain once some part of the grain starts crystallizing. It is likely that crystalline ice grains first start populating at the outer radii of the plasma where gas is slightly warmer, while amorphous ice grains form near the electrodes. The vortex motion observed in Figure \ref{cloud photos} then transports grains throughout the plasma, mixing the originally-stratified amorphous and crystalline regions. The phase transition from amorphous to crystalline ice is irreversible, a well-known phenomenon previously observed in ice film measurements \citep{gudipati2023thermal}, so amorphous grains that are transported to warmer areas may undergo the irreversible phase transition and become crystalline.

\begin{figure*}
  \centering
  \includegraphics[width=15cm]{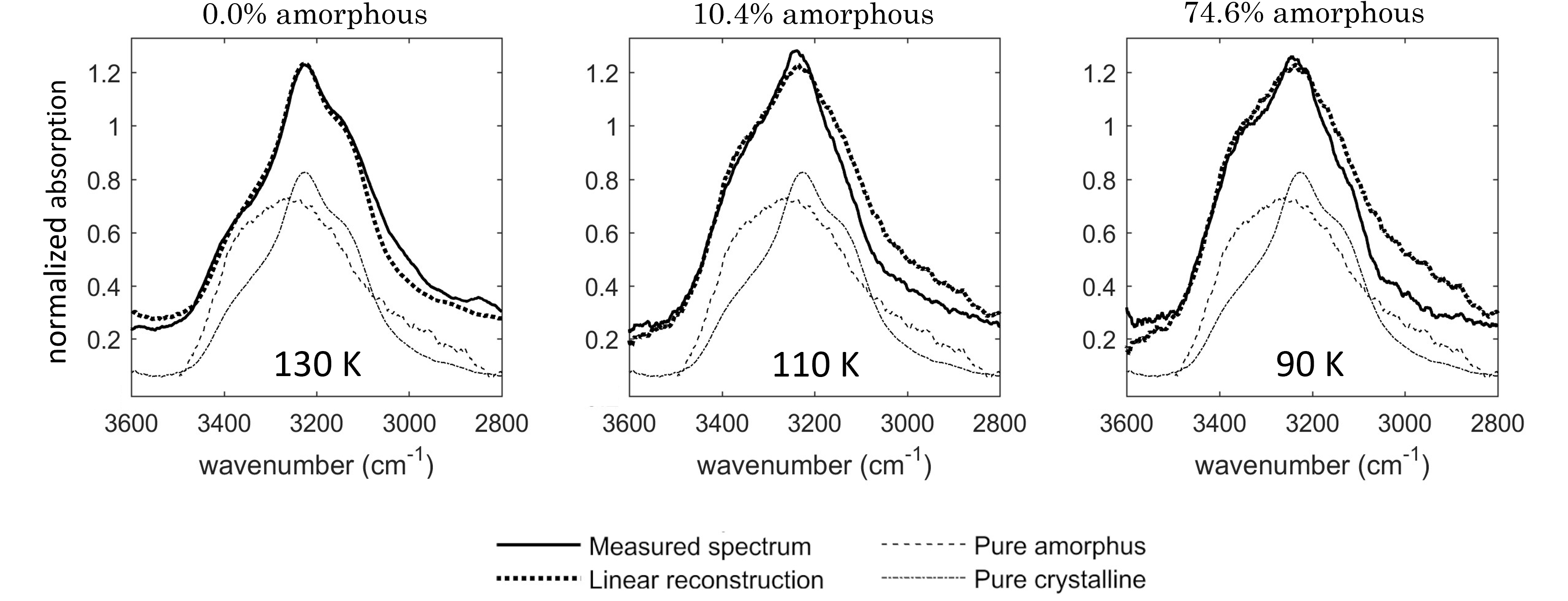}
  \caption{Spectra at 130, 110, and 90 K, in 400 mTorr of Hydrogen plasma (solid line), compared to the fitted linear combination of pure amorphous and crystalline phases (dotted line). Thin dashed and dot-dashed lines show the pure amorphous and crystalline spectra (shifted downwards by 0.2 for clarity). Table \ref{amorphous table} gives percentages for all tested pressures and temperatures.}
  \label{superpositions}
 \end{figure*}

To determine the fraction of ice grains in each phase, a simple linear mixing model is devised; because grains are not densely packed, nonlinear mixing effects should be negligible \citep{nonlinearmixing}. Spectral mixing to quantify amorphous and crystalline ice has been successfully used in previous works \citep{fama2010,berdis2020}, and this analysis is in line with those previous works. The measured spectrum is fit to a linear combination of the pure-amorphous and pure-crystalline spectra, minimizing the expression:
\begin{equation}
    ( a A_{a} + (1-a) A_{c} - A_{tot})^2,
    \label{fit eqn}
\end{equation}

\noindent where $A_{tot}$ is the measured signal, $A_{a}$ is the pure-amorphous spectrum, and $A_{c}$ is the pure-crystalline spectrum. This ignores the change in the crystalline absorption peak location with temperature, but that change is small compared to the crystalline-to-amorphous transition. The parameter $a$ is constrained within the range [0,1], representing the fraction of grains in the pure-amorphous state. All spectra were normalized before the fit. The fits for 130, 110, and 90 K at 800 mTorr are presented in Figure \ref{superpositions}. Results for all $t = 2$-min scans are given in Table \ref{amorphous table}. 

As is expected, the fraction of amorphous ice increases as temperature decreases. At 800 mTorr, no amorphous ice is observed above 130 K, nearly all ice is amorphous at 80 K, and intermediate temperatures show mixed compositions. Because amorphous ice is known to exist below 130 K \citep{mastrapa2013amorphous}, we conclude the warmest parts of the ice cloud are roughly 50 K above the electrode temperature, such that they are 130 K when electrodes are 80 K; the coldest ice grains are within 10 K of electrode temperature, as amorphous ice is seen at 120 K. Pressure greatly affects this: at 1600 mTorr, 33.1\% of the ice is amorphous at 120 K while there is little to none at lower pressures, while at 400 mTorr, there is still some crystalline ice at 80 K. This variation is likely because the cold background gas is less collisional at low pressure, and thus cools the incoming water vapor less efficiently. This also explains why grain size does not correlate with temperature at 400 mTorr as reliably as it does at higher pressures: the temperature profile is less consistent within the icy region. It is hypothesized that if the water vapor were cooled before entering the chamber, this effect would disappear, as vapor would not require as many collisions to equilibrate with the cold background gas.

\begin{table}
  \centering
  \caption{Percentage of amorphous ice with pressure and temperature, derived from FTIR spectra taken immediately after closing H$_2$O inlet valve at $t=2$ min. At all temperatures above 130 K, only crystalline ice was present.}
  \label{amorphous table}
     \begin{tabular}{ c | c c c c c c}
     \hline 
     \hline
      & 130 K & 120 K & 110 K & 100 K & 90 K & 80 K\\
     \hline
     0.4 Torr & 0\% & 0\% & 10.4\% & 64.1\% & 74.6\% &  60.5\%\\
     0.8 Torr & 0\% & 1.7\% & 89.4\% & 73.4\% & 83.0\% & 94.5\% \\
     1.6 Torr & 0\% & 33.1\% & 51.1\% & 76.1\% & 100\% & 100\% \\
     \hline 
     \hline
     \end{tabular}
\end{table}

\begin{figure*}
  \centering
  \includegraphics[width=15cm]{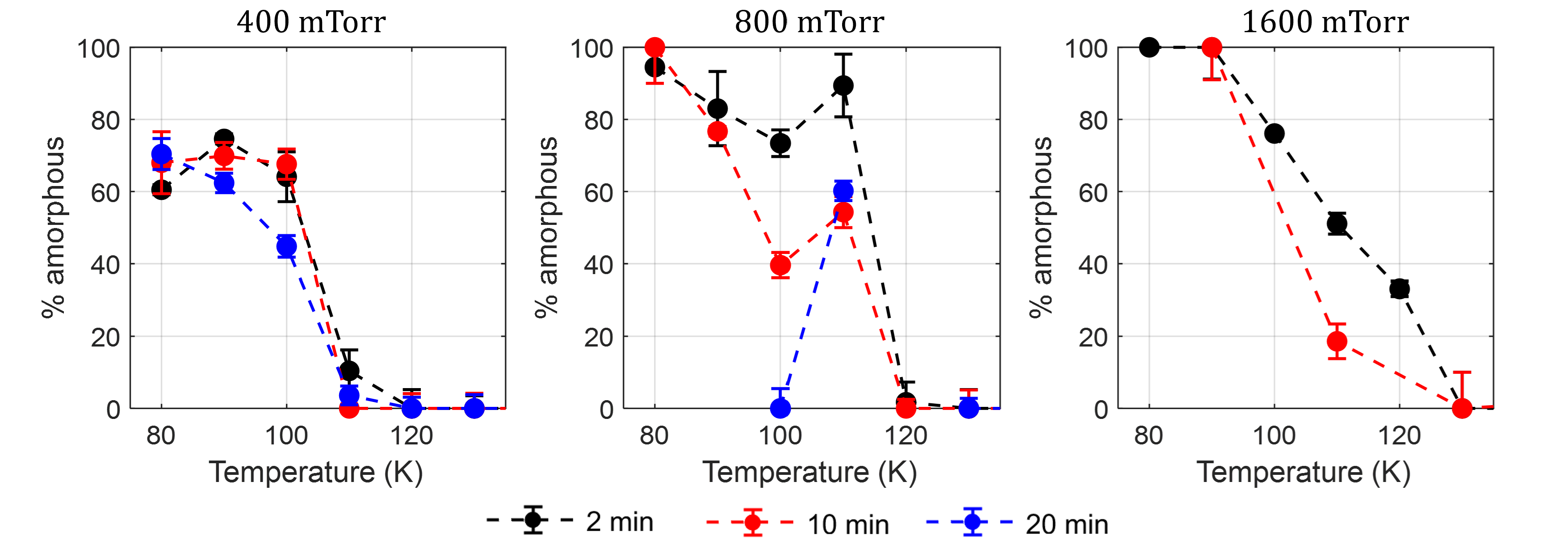}
  \caption{Percentage of ice in the amorphous phase in each FTIR scan. Each of the three plots shows the time-evolution of phase composition at five temperatures for a single pressure. Error bars give the root mean square (RMS) deviation of the fit to of the linear mixing model to the data. Gaps in data are due to the disappearance of ice through sublimation.}
  \label{percent amorphous fig}
 \end{figure*}

This method for determining phase composition is only reliable for the $t=2$ minute FTIR spectrum of each configuration. As grains grow, inelastic scattering alters the spectra and causes the fitting procedure to fail. This can be accounted for by first adjusting the pure-amorphous and pure-crystalline extinction coefficients to include Mie scattering using the effective radii given in Fig. \ref{grain sizes}. We can then perform the fit in Eq. \ref{fit eqn} to obtain the percent of grains in each phase as before. The results are presented in Figure \ref{percent amorphous fig}.

The ice generally becomes more crystalline over time. For temperatures above 90 K, the fraction of amorphous grains was lower at $t = 10$ minutes than at 2 minutes, and generally lower still at 20 minutes. Amorphous ice naturally crystallizes on a timescale that is greatly dependent on its temperature: while ice at 90 K takes about 3000 years for crystallization to begin, ice at 130 K takes only about an hour until onset of crystallization \citep{mastrapa2013amorphous}. If ice is some tens of kelvins above the temperature of the electrodes, it is plausible that some amorphous ice, held slightly above 130 K, could begin to crystallize in the 20 minutes spanned by the experimental sequence. It must also exist at a low enough temperature to not transition so fast as to appear crystalline already in the 2-minute FTIR scan. The vortex motion of ice grains within the cloud could accelerate the phase transition onset while slowing its course: as ice grains circulate between the colder and warmer regions of the plasma, they transiently heat and can begin to crystallize before quenching again. Collisions with non-equilibrium gas molecules or ion bombardment into the negatively-charged ice surface could also heat the ice and instigate crystallization.

\section{Observations of ice nucleation}

An experiment was conducted in which vapor was introduced to the chamber under the same conditions as previously tested, but without the RF power engaged. If ice grains formed through heterogeneous nucleation due to aerosolized impurities, one would expect nucleation to occur; however, no ice grains were observed, with ice instead forming a film on the surface of the electrodes. This indicates that the presence of plasma is fundamental to the nucleation of ice grains in this experiment, and that impurities are not numerous enough to seed ice nucleation on their own. 

Several previous experiments \citep{bouchoule1999dusty, shukla_mamun_2002} have used similar capacitively-coupled RF plasmas, but instead placed prefabricated plastic microspheres on the bottom electrode, and observed that these particles acquire charge and levitate into the plasma through the same mechanism that confines ice grains in this study. An analogous process has been proposed for the lunar surface, in which a flux of solar wind and ultraviolet radiation causes dust to become charged and levitate into the weakly-ionized lunar atmosphere \citep{lisin2015lunar}. Considering these results, it has been suggested that ice could form in microscopic patches on the cold electrode surface, acquire charge, and levitate into the plasma, providing an ice nucleus around which vapor could accrete. However, this hypothesis is not supported by observation in the experiments reported here. Formation of ice on surfaces becomes more efficient at lower temperature, while as shown in Figures \ref{cloud photos} and \ref{magnitudes}, less ice was observed in the plasma at cold temperatures. We carried out control experiments in which an ice film was first deposited on the electrodes and RF power was then applied. No ice grains were observed in the plasma; all ice remained on the electrodes. In another experiment, when ice grains were first formed in the plasma, the RF source was subsequently shut off so that grains fell to the electrodes, and then the RF power was restored, grains remained stuck to the electrodes and did not levitate back into the plasma. This suggests that ice grains do not nucleate on the electrode surface and then levitate, but instead nucleate directly within the plasma.

This indicates that homogeneous nucleation is the most probable mechanism for ice grain formation. Gas-phase ice nucleation has been traditionally disregarded as inefficient based on arguments from classical nucleation theory (CNT). However, the validity of CNT has been questioned for application to ice nucleation in this context, as it requires extrapolating surface tension from liquid to the solid phase \citep{makkonen2012misinterpretation}, assumes thermodynamic equilibrium when it does not apply \citep{rapp2006modeling}, and does not consider the molecular mechanics of the nucleation process \citep{bellan2022}. An efficient alternative to the nucleation described by CNT is ion-catalyzed nucleation, in which ions attract the dipole moment of water molecules to form a thermodynamically stable nuclei, onto which more vapor can condense. This has been studied in the context of ice nucleation in the Earth’s mesosphere \citep{witt1969, arnold1980}, and is the underlying mechanism of condensation in cloud chambers, as first noted by \citet{wilson1897}: a high-energy particle (e.g. an x-ray or alpha particle) ionizes supersaturated vapor and instigates condensation around the ion \citep{adachi1992}. Similarly, high-energy electrons accelerated by the RF field in this experiment ionize both the background gas and the water vapor, and these ions could seed ice nucleation. 

A study by \citet{bellan2022} investigated this mechanism for the conditions of this experiment, and showed OH$^-$ ions created by dissociation of water vapor by energetic electron impacts were the best candidates to attract water molecules and efficiently form OH$^-$(H$_2$O)$_n$ clusters that then serve as nuclei for ice grain production. This mechanism predicts that nucleation will cease at low pressure and at high RF power, in which electrons are accelerated by the RF electric field to energies above the electron affinity of OH$^-$, and knock electrons off the negative ions before ice nuclei can form around them; this agrees with observations reported by \citet{chai2015} in their similar cryogenic plasma experiment. \citet{chai2015} were also able to form ice grains with other polar molecules (acetone, methanol), but were unable to form ice grains from non-polar molecules (CO$_2$), suggesting that the dipole moment is instrumental in the nucleation process, a prediction of the ion-catalyst theory by \citet{bellan2022}.

In this study, all experimental configurations used identical mass flow rates of water vapor injected during the first two minutes of each run, and no water vapor was injected after 2 minutes. The same amount of vapor therefore passes through the chamber in each run, though the vapor density varies with pressure and temperature. Some fraction of the injected water vapor is contained in the ice grains, while the remainder either freezes to the electrodes or is pumped out of the chamber. As less ice was detected in the plasma at lower temperatures, both through IR absorption and through direct imaging, the remainder of the injected vapor likely condensed on the electrodes. A thin film of ice was indeed visible on the electrodes at temperatures of 90 K and below, while this was not observed at warmer temperatures. This suggests that at colder temperatures, water efficiently freezes on surfaces in addition to forming small grains through homogeneous nucleation. At higher temperatures, homogeneous nucleation is very efficient at producing large crystalline grains, and any ice condensed on the electrodes sublimes and accretes onto the ice grains in the plasma.

\section{Conclusions and astrophysical implications}
In summary, micron-scale grains of pure H$_2$O ice are observed to form spontaneously when water vapor is injected into a cryogenically-cooled radio-frequency plasma discharge. These grains have an elongated fractal-like structure. Both amorphous and crystalline (Ih) phases of ice are observed. Though the amorphous and crystalline phases have significantly different IR absorption spectra as expected, no significant difference in ice morphology is observed. Grains that are purely-crystalline and purely-amorphous can coexist within the plasma, either due to nucleating at different temperatures, or upon undergoing an irreversible amorphous-to-crystalline phase transition; this causes ice contained in the plasma to become more crystalline over time, potentially due to heating and cooling as ice grains traverse the non-uniform temperature profile of the plasma. At lower electrode temperatures, amorphous ice takes longer to transition to the crystalline phase. Grains increase in size over time, observed both by direct imaging and by interpretation of Mie scattering of the infrared spectrum. Based on observations of nucleation at different temperatures and characteristics of the ice’s infrared absorption spectrum, along with measures taken to remove impurities of the system, it is suggested that ice nucleates homogeneously around negative ions formed by the dissociation of water by high-energy electron impacts within the plasma.

The experimental setup deviates from conditions in space in terms of pressure, temperature, and ionization fraction. The temperature range in the experiment aligns with that found in protoplanetary disks but exceeds that of interstellar clouds. Pressure is significantly higher in the experiment than in either astrophysical situation. This leads to a large difference in density: neutral densities in the experiment are on the order of $10^{16}$ cm$^{-3}$, while the densest parts of protoplanetary disks are smaller by a factor of $10^2$, and molecular clouds by a factor of up to $10^{10}$ \citep{nicolov2023}. Despite this low density, essential processes governing ice formation, growth, and morphology should be similar. Although the coalescence of water molecules into ice grains is slower due to the lower collision frequencies and larger inter-molecular spacing, ice can still nucleate and grow, since sublimation is also comparatively weaker at these pressures and temperatures. Both the nucleation mechanism \citep{bellan2022} and growth process \citep{bellan2020} rely on ice gaining a sufficient amount of charge through electron collection, a condition satisfied in both protoplanetary and interstellar environments \citep{nicolov2023}. While the rates of observed processes will differ from those in the laboratory, and further theoretical work must be done to establish the form of this scaling, the fundamental mechanisms should remain valid.

This research observed that the ice grains that formed had a spindle-like fractal morphology; in this experiment, the total grain length stretched to over ten times the characteristic size obtained from a simple Mie scattering model of the absorption spectrum. As properties of astrophysical ice and dust grains are typically inferred through such radiative transfer models of infrared absorption measurements, the true size of these grains may be underestimated if they have similar structures to those observed in the laboratory, which are far removed from the spherical grain approximation used for convenience in modeling studies. As grains in this study grew larger and more rapidly at lower pressures, the even-lower pressures in protoplanetary and interstellar conditions may lead to formation of significantly larger ice structures. Conversely, smaller grains were observed to grow in colder temperatures. 

The ability of ice grains to self-nucleate from cold water vapor under weakly ionized plasma conditions carries profound implications for ice grain formation in interstellar and protoplanetary environments. This research demonstrates that both metastable amorphous and stable crystalline ice grains can arise through this homogeneous nucleation process, depending on the prevailing thermal conditions. 

Based on these findings, we propose that in the outer reaches of protoplanetary disks where temperatures fall well below 130 K, smaller amorphous ice grains tend to form, while larger crystalline grains form in the warmer conditions further inward. This creates a gradient of grain size and ice phase with increasing distance from the ice line. Nucleation will cease at an inner radius, at which either neutral temperatures are too high, or electrons are sufficiently energetic to neutralize negative ions before they can attract water molecules and grow to stable nuclei. In contrast, within dense molecular clouds where temperatures are even lower, we anticipate that homogeneous nucleation can form small amorphous ice grains, even in the absence of the refractory dust typically understood to be necessary for nucleation. Based on our experimental observations that ice tended to form more efficiently on the electrode surfaces at the coldest temperatures tested, at the expense of ice accreting to grains in the plasma, it is likely that heterogeneous ice nucleation becomes more common in the colder interstellar conditions when an abundance of refractory dust is present; however, we expect ion-catalyzed homogeneous nucleation to occur in parallel with heterogeneous nucleation in these weakly-ionized plasma environments.
\\

Acknowledgements: Experimental work is supported by the NSF/DOE Partnership in Basic Plasma Science and Engineering via USDOE Award DE-SC0020079. The contribution from M. S. Gudipati was supported through funding from NASA DDAP and JPL's JROC programs and was carried out at the Jet Propulsion Laboratory, California Institute of Technology, under a contract with the National Aeronautics and Space Administration (80NM0018D0004). We thank Dr. Zhehui Wang at Los Alamos National Laboratory for the loan of a residual gas analyzer, obtained with support from USDOE Fusion Energy Sciences.

\bibliographystyle{aasjournal}
\bibliography{bibliography.bib}
\end{document}